\documentclass[aps,twocolumn, noshowpacs, floatfix]{revtex4}
\usepackage{amsfonts}
\usepackage{amssymb}
\usepackage{graphicx}
\usepackage{amsmath}
\usepackage[english]{babel}
\usepackage{color}

\begin{document}

\title{Polaron dynamics with a multitude of Davydov D$_2$ trial states}

\author{Nengji Zhou$^{1,2}$, Zhongkai Huang$^{1}$, Jiangfeng Zhu$^{1}$, Vladimir Chernyak$^{1,3}$, and Yang Zhao$^{1}$\footnote{Electronic address:~\url{YZhao@ntu.edu.sg}}}
\date{\today}
\affiliation{$^1$Division of Materials Science, Nanyang Technological University, Singapore 639798, Singapore\\
$^2$Department of Physics, Hangzhou Normal University, Hangzhou 310046, China \\
$^3$Department of Chemistry, Wayne State University, Detroit, USA}

\begin{abstract}
We propose an extension to the Davydov D$_2$ {\it Ansatz} in the dynamics study of the Holstein molecular crystal model with diagonal and off-diagonal exciton-phonon coupling using the Dirac-Frenkel time-dependent variational principle. The new trial state by the name of the ``multi-D$_2$ {\it Ansatz}" is a linear combination of Davydov D$_2$ trial states, and its validity  is carefully examined by quantifying how faithfully it follows the Schr\"odinger equation. Considerable improvements in accuracy have been demonstrated in comparison with the usual Davydov trial states, i.e., the single D$_1$ and D$_2$ {\it Ans\"{a}tze}. With an increase in the number of the Davydov D$_2$ trial states in the multi-D$_2$ {\it Ansatz}, deviation from the exact Schr\"odinger dynamics is gradually diminished, leading to a numerically exact solution to the Schr\"odinger equation.
\end{abstract}

\maketitle

\section{Introduction}

\begin{figure}[tbp]
\centering
\includegraphics[scale=0.4]{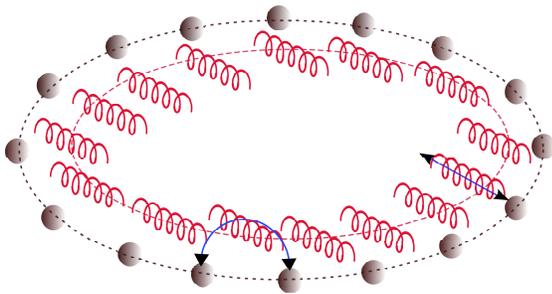}
\caption[FIG]{ Schematic of the Holstein model. In the molecular ring with $N = 16$ sites, the circles and oscillators denote
the excitons and phonons, respectively, and the arrows represent the diagonal coupling (one on one) and off-diagonal coupling (one on two)
between the excitons and phonons.  }
\label{polaron}
\end{figure}

Since the advent of the ultrafast laser spectroscopy,
much attention has been devoted to the
relaxation dynamics of photoexcited entities, such as polarons
in inorganic liquids and solids \cite{an_04, zheng_06, liu_06}, charge carriers in topological
insulators \cite{bron_14, tim_12}, electron and hole trapping of semiconductor nanoparticles \cite{chi_02, kli_00,kim_96},
and electron-hole pairs in light-harvesting complexes of photosynthetic organisms \cite{sau_79, reng_01, blan_02, gron_06}. In photosynthesis, it is suggested that
quantum coherence might play a significant role
in achieving the remarkably high efficiency ($>95\%$) of the excitation energy transport \cite{eng_10, coll_10, pan_10}, which can be studied
by observing electronic superpositions and their evolution using $2$D electronic spectroscopy
techniques \cite{eng_10, coll_10, coll_09, cal_11}. Emerging technological capabilities to control femtosecond
pulse durations and down-to-one-hertz bandwidth resolutions
provide novel probes on vibrational dynamics and excitation relaxation
which were elusive in the past \cite{tom_00}. Recently, developments in
ultrafast laser physics and technologies allow us to study the
nonequilibrium carrier/exciton dynamics that was previously
inaccessible to traditional linear optical spectroscopy.
In contrast, modeling of polaron dynamics has not
received much deserved attention.

The Holstein molecular crystal model has been extensively used to study properties of polarons in molecular crystals and biological systems \cite{hol_59}. As an example, a molecular ring with $16$ sites, each coupled with a phonon mode, is shown in Fig.~\ref{polaron}. Two kinds of exciton-phonon interactions can be included in the Holstein model, namely, the diagonal coupling (marked by an arrow attached to a single site) as a nontrivial dependence of the exciton site energies on the lattice coordinates, and the off-diagonal coupling (marked by an arrow attached to two nearest-neighboring sites) as a nontrivial dependence of the exciton transfer integral on the lattice coordinates \cite{su_79}. Simultaneous presence of diagonal and off-diagonal coupling seems crucial to characterize solid-state excimers, where
a variety of experimental and theoretical considerations imply a strong dependence of electronic tunneling upon certain coordinated distortions of neighboring molecules in the formation of bound excited states \cite{dis_84, sumi_89}. However, in the literature, little attention has been paid to the Hamiltonians containing the off-diagonal exciton-phonon coupling due to inherent difficulties to obtain
reliable solutions \cite{mahan_00}, especially for the polaron dynamics \cite{zhao_12}. Early treatments of off-diagonal coupling
include the Munn-Silbey theory \cite{zhao_94, dmchen_11} which is based upon a perturbative approach with added constraints on canonical
transformation coefficients determined by a self-consistency equation. The global-local (GL){\it Ansatz} \cite{zhao_94b, zhao_08}, formulated
by Zhao {\it et al.}~in the early $1990$s, was later employed in combination with the dynamic coherent potential approximation (with
the Hartree approximation) to arrive at a state-of-the-art ground-state wave function as well as higher eigenstates \cite{liu_09}.

In the absence of an exact solution, various numerical approaches were developed in the past few decades, including the exact diagonalization (ED) \cite{alex_94, mell_97}, quantum Monte Carlo (QMC) simulation \cite{rae_83, wang_89, kor_98}, variational method \cite{cata_99, tana_03, per_04, cata_04, bar_07, zhao_08}, density matrix renormalization group (DMRG) \cite{whit_92, jeck_98}, the variational exact diagonalization (VED) \cite{wei_00, ku_07}, and the method of relevant coherent states \cite{bar_07}. Most of these approaches were designed to probe the ground-state properties. For excited-state properties and dynamics of the polaronic systems, however, few of them provide a satisfactory resolution. For example, a time-dependent variant of DMRG, i.e., t-DMRG \cite{white_04, fei_05, zhao_08b}, was developed to elucidate the polaron dynamics. Yet, it cannot accurately simulate the system dynamics from an arbitrary initial state, since high-lying excited states
can not be adequately described by DMRG. Fortunately, the variational approach is still effective in dealing with polaron dynamics so long as a proper trial wave function is chosen. Previously, static properties of the Holstein polaron have been examined using a series of trial wave
functions based upon phonon coherent states, such as the Toyozawa {\it Ansatz} \cite{zhao_94b, meier_97,zhao_97}, the GL {\it Ansatz} \cite{zhao_94b, zhao_97, zhao_08, zhao_97b}, and a delocalized form of the Davydov $\rm D_1$ {\it Ansatz} \cite{sun_13}. By using these {\it Ans\"atze}, the ground state band and the self-trapping phenomenon were adequately investigated.
To simulate the time evolution of the Holstein polaron, at least two Davydov Ans\"{a}tze,
namely, the ${\rm D_1}$ and ${\rm D_2}$ Ans\"{a}tze \cite{skri_88, for_93, han_94},
were used following the Dirac-Frenkel variation scheme \cite{dira_30},
a powerful apparatus to reveal accurate dynamics of quantum
many-body systems. Time-dependent variational parameters
which specify the trial state are obtained from solving
a set of coupled differential equations generated by the
Lagrangian formalism of the Dirac Frenkel variation.
Validity of this approach is carefully checked by quantifying
how faithfully the trial state follows the Schr\"odinger equation \cite{luo_10, sun_10, zhao_12}.
Numerical results show that the ${\rm D_1}$ {\it Ansatz} is
effective and accurate in studying the Holstein polaron dynamics with the diagonal coupling, but fails to describe the
case with the off-diagonal coupling. Despite being a simplified version of the $\rm D_1$ trial state,
however, the ${\rm D_2}$ {\it Ansatz} instead can deal with the off-diagonal coupling case,
albeit with non-negligible deviation from the exact solution to the Schr\"odinger dynamics \cite{zhao_12}. The purpose of this paper is to test the feasibility of using a superposition of the Davydov trial states
to study the dynamics of the Holstein model with simultaneous diagonal and off-diagonal coupling.
For simplicity, only the multi-${\rm D_2}$ {\it Ansatz} is probed in this work, and the validity of the trial state will be comprehensively
investigated.

The paper is organized as follows. In Sec. II we introduce
the model Hamiltonian and the multi-${\rm D_2}$ {\it Ansatz} for the studies of the polaron dynamics.
The quantitative measure of the validity of the variational method is explained.
In Sec. III, numerical results from our investigation on the dynamics of the Holstein
polaron in the diagonal and off-diagonal coupling regimes are displayed and discussed.
Conclusions are drawn in Sec.~IV.

\section{Methodology}

The one-dimensional Holstein molecular crystal model for the exciton-phonon system can be described by the Hamiltonian below
\begin{equation}
\hat{H} = \hat{H}_{\rm ex} + \hat{H}_{\rm ph} + \hat{H}_{\rm ex-ph}^{\rm diag} + \hat{H}_{\rm ex-ph}^{\rm o.d.},
\label{holstein}
\end{equation}
where $\hat{H}_{\rm ex}, \hat{H}_{\rm ph}$ and $\hat{H}_{\rm ex-ph}$ correspond to the exciton Hamiltonian, bath (phonon) Hamiltonian and exciton-phonon coupling Hamiltonian
defined as
\begin{eqnarray}
\hat{H}_{\rm ex} & = & -J \sum_{n}\hat{a}_n^{\dag}(\hat{a}_{n+1} + \hat{a}_{n-1}), \nonumber \\
\hat{H}_{\rm ph} & = & \sum_{q} \omega_{q}\hat{b}_{q}^{\dag}\hat{b}_{q},  \nonumber \\
\hat{H}_{\rm ex-ph}^{\rm diag} & = & -g \sum_{n,q} \omega_q \hat{a}_n^{\dag}\hat{a}_n(e^{iqn}\hat{b}_q+e^{-iqn}\hat{b}_q^{\dagger}), \nonumber \\
\hat{H}_{\rm ex-ph}^{\rm o.d.} & = &  \frac{1}{2}\phi \sum_{n,q}\omega_q \left\{ \hat{a}_n^{\dag}\hat{a}_{n+1}[e^{iqn}(e^{iq} - 1)\hat{b}_q + {\rm H.c.}] \right. \nonumber \\                  && \left. + \hat{a}_n^{\dag}\hat{a}_{n-1} [ e^{iqn}(1 - e^{-iq})\hat{b}_q + {\rm H.c.} ] \right\},
\label{Hamiltonian}
\end{eqnarray}
where $\rm H.c.$ denotes the Hermitian conjugate, $\omega_q$ is the phonon frequency at the momentum $q$, $\hat{a}_n^{\dag}$ ($\hat{a}_n$) is the exciton creation (annihilation) operator
for the $n$-th molecule, and $\hat{b}_q^{\dag}$ ($\hat{b}_q$) is the creation (annihilation) operator of a phonon with the momentum $q$,
\begin{equation}
\hat{b}_q^{\dag} = N^{-1/2}\sum_n e^{iqn}\hat{b}_n^{\dag}, \quad \hat{b}_n^{\dag} =  N^{-1/2}\sum_q e^{-iqn}\hat{b}_q^{\dag}.
\label{momentum}
\end{equation}
The parameters $J, g$ and $\phi$ represent the transfer integral, diagonal and off-diagonal coupling strengths, respectively, and $N=16$ is the number of sites in the Holstein polaron.
In this paper, a linear dispersion phonon band is assumed,
\begin{equation}
\omega_q = \omega_0\left.[1 + W(2|q|/\pi)-1\right.],
\label{dispersion}
\end{equation}
where $\omega_0$ denotes the central energy of the phonon band, $W$ is the band width between $0$ and $1$, and the momentum is set to be
$q=2\pi l/N$ with $(l=-\frac{N}{2}+1, \ldots, \frac{N}{2})$.

A trial state, termed as the ``Davydov multi-$\rm{D}_2$ {\it{\it Ansatz}}," is adopted
\begin{eqnarray} \label{wave}
&& |{\rm D}^M_2(t)\rangle = \\ \nonumber
&& \sum_{n=1}^N \hat{a}_n^{\dag}|0\rangle_{\rm ex}\sum_{i=1}^M \psi_{i,n}(t)\exp\left[\sum_q\left(\lambda_{i,q}\hat{b}_q^{\dag}-\lambda_{i,q}^*\hat{b}_q\right)\right] |0\rangle_{\rm ph},
\end{eqnarray}
where $\hat{a}_n^{\dag}(\hat{a}_n)$ is the creation (annihilation) operator of a exciton at the $n$-th site, $\hat{b}_q^{\dag}(\hat{b}_q)$ is the creation (annihilation) operator of a phonon with momentum $q$, and the variational parameters $\psi_{i,n}$ and $\lambda_{i,q}$ denote the exciton probability  and phonon displacement, respectively.
Moreover, $n$ and $i$ represent the ranks of the site in the molecular ring and the coherent superposition state, respectively.

The equations of the motion are derived for the variational parameters $\psi_{i,n}$ and $\lambda_{i,q}$ by adopting the Dirac-Frenkel variational method, in which the Lagrangian $L$ is formulated as
\begin{eqnarray}
L & = & \langle {\rm D}^M_2(t)|\frac{i\hbar}{2}\frac{\overleftrightarrow{\partial}}{\partial t}- \hat{H}|{\rm D}^M_2(t)\rangle \nonumber \\
& = & \frac{i\hbar}{2}\left[ \langle {\rm D}^M_2(t)|\frac{\overrightarrow{\partial}}{\partial t}|{\rm D}^M_2(t)\rangle - \langle {\rm D}^M_2(t)|\frac{\overleftarrow{\partial}}{\partial t}|{\rm D}^M_2(t)\rangle \right] \nonumber \\
&-& \langle {\rm D}^M_2(t)|\hat{H}|{\rm D}^M_2(t)\rangle.
\label{Lagrangian}
\end{eqnarray}
From this Lagrangian, the equation of the motion for $\alpha$ and its time derivative $\dot{\alpha}(t)$ can be obtained,
\begin{equation}
\frac{d}{dt}\left( \frac{\partial L}{\partial \dot{\alpha}^*}\right) - \frac{\partial L}{\partial \alpha^*} = 0,
\label{motion}
\end{equation}
where $\alpha$ is one of the variational parameters $\psi_{i,n}$ and $\lambda_{i,q}$ in Eq.~(\ref{wave}).
Details on derivation of the equations of the motion for the Holstein polaron dynamics with a multi-$\rm{D}_2$ {\it Ansatz}
are given in Appendix~A.

\begin{figure}[tbp]
\centering
\includegraphics[scale=0.4]{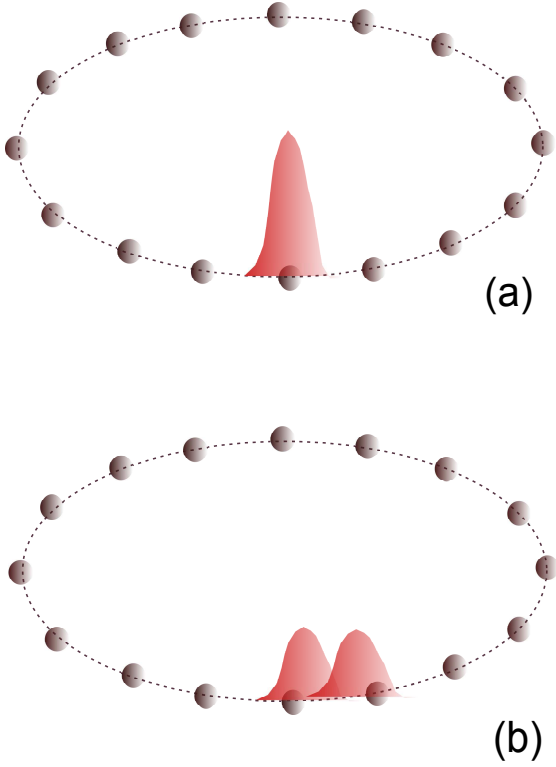}
\caption[FIG]{ Schematics of the initial states of the Holstein polaron are given in (a) and (b) for the diagonal and off-diagonal coupling cases, respectively.
The exciton is prepared in (a) only at the site $n=N/2$, i.e., $\psi_n(t=0)=\delta_{n, N/2}$, while in (b) the exciton is created at the two nearest neighboring sites, i.e., $\psi_{n}=(\delta_{n,N/2}+\delta_{n,N/2+1})/\sqrt{2}$. The phonon displacement coefficient $\lambda_n(t=0)=0$ is set.}
\label{init}
\end{figure}

As shown in Fig.~\ref{init}, the initial states of the Holstein polaron are prepared for the diagonal coupling case with the coupling strength $g>\phi=0$ in (a) and off-diagonal coupling case with the coupling strength $\phi>g=0$ in (b). In order to avoid the singularity, a little noise uniformly distributed from $[-\varepsilon, \varepsilon]$ is added with $\varepsilon = 10^{-5}$ in the initial state for both $\psi_{i,n}$ and $\lambda_{i,q}$. For each set of the coefficients $W, g, J$ and $\phi$ defined in Eqs.~(\ref{Hamiltonian}) and (\ref{dispersion}), more than $100$ initial states are used in the simulations. A power-law relation between the CPU time and the multiplicity $M$ has been found in our time-dependent variational approach,
and the value of the exponent $2.8$ indicates that the time complexity of the program is $O(n^3)$. Even for the largest value of the multiplicity in our paper, i.e., $M=64$, the memory usage of the multi-$\rm D_2$ {\it Ansatz} is $256$ MB (million bytes), still bearable for the computation. Though both the CPU time and  memory usage of the multi-$\rm D_2$ {\it Ansatz} are much larger than those of the single $D_2$ {\it Ansatz} (less than $1$ hour for the CPU time and $0.1$ MB for the memory usage), the multi-$D_2$ {\it Ansatz} has substantially improved the accuracy on variational dynamics. The energy of the system $E_{\rm total}=E_{\rm ex}+E_{\rm ph}+E_{\rm diag} + E_{\rm off}$ is calculated based on the multi-$\rm{D}2$ {\it Ansatz} in Eq.~(\ref{wave}),
\begin{eqnarray}
E_{\rm ex} &=& \langle {\rm D}^M_2| \hat{H}_{\rm ex}|{\rm D}^M_2\rangle \nonumber \\
           &=&-J\sum_{i,j}^{M}\sum_{n}\psi_{j,n}^*(\psi_{i,n+1}+\psi_{i,n-1})S_{j,i}, \nonumber \\
E_{\rm ph} &=& \langle {\rm D}^M_2| \hat{H}_{\rm ph}|{\rm D}^M_2\rangle \nonumber \\
           &=&\sum_{i,j}^{M}\sum_{n}\psi_{j,n}^*\psi_{i,n}\sum_q\omega_q\lambda_{j,q}^*\lambda_{i,q}S_{j,i},  \\
E_{\rm diag} &=& \langle {\rm D}^M_2| \hat{H}_{\rm ex-ph}^{\rm diag}|{\rm D}^M_2\rangle = -g \sum_{i,j}^{M}\sum_n\psi_{j,n}^*\psi_{i,n} \nonumber \\
             &&\sum_q\omega_q(e^{iqn}\lambda_{i,q} +e^{-iqn}\lambda_{j,q}^*)S_{j,i}, \nonumber \\
E_{\rm off} &=&  \langle {\rm D}^M_2| \hat{H}_{\rm ex-ph}^{\rm o.d.}|{\rm D}^M_2\rangle = \frac{1}{2}\phi \sum_{i,j}^{M}\sum_n \omega_q S_{j,i} \nonumber \\
            && \left \{\psi_{j,n}^*\psi_{i,n+1}[e^{iqn}(e^{iq} - 1)\lambda_{i,q} + {\rm H.c.}] \right. \nonumber \\
            && \left. + \psi_{j,n}^*\psi_{i,n-1}[ e^{iqn}(1 - e^{-iq})\lambda_{i,q} + {\rm H.c.} ] \right\}, \nonumber
\label{energy}
\end{eqnarray}
where $S_{j,i}=\langle \lambda_j|\lambda_i\rangle$ is the Debye-Waller factor defined as
\begin{equation}
S_{j,i}= \exp\left\{\sum_q\lambda_{j,q}^*\lambda_{i,q} -\frac{1}{2}(|\lambda_{i,q}|^2+|\lambda_{j,q}|^2) \right\}.
\label{factor}
\end{equation}
The normalization of the wave function $Norm = \langle {\rm D}^M_2|{\rm D}^M_2\rangle$ is also calculated for conservation.

Furthermore, the exciton probability  $P_{\rm ex}(t, n)$ and the phonon displacement $X_{\rm ph}(t, n)$ are also
calculated for the dynamics of the Holstein polaron.  Firstly, the reduced single-exciton density matrix $\rho_{mn}(t)=\rm{Tr}[\rho(t)\hat{a}_m^\dag\hat{a}_n]$ is obtained by solving the coupled equations of variational parameters, where $\rho(t)=|{\rm D}^M_2\rangle\langle {\rm D}^M_2|$ is the full density matrix at the zero temperature. After substituting
the trial state $|{\rm D}^M_2(t)\rangle$ of Eq.~(\ref{wave}), the reduced density matrix is then derived as
\begin{equation}
\rho_{mn}(t)= \sum_{i,j}^M\psi_{j,m}^{*}\psi_{i,n}S_{j,i}.
\label{density}
\end{equation}
Thus, the exciton probabilities $P_{\rm ex}(t, n)=\rho_{nn}(t)$ ($n=1,2,\ldots, N$) are obtained from the diagonal elements of the reduced density matrix.
The phonon displacement $X_{\rm ph}(t, n')$ in real space at the ${n'}-$th site is calculated by
\begin{eqnarray}
X_{\rm ph}(t, n') & = & \langle {\rm D}^M_2|b_{n'}(t)+ b_{n'}(t)^{\dag}|{\rm D}^M_2\rangle   \\
       & = & \frac{1}{\sqrt{N}}\sum_qe^{iqn'}\left[\sum_{i,j}\sum_{n}\lambda_{i,q}\psi_{j,n}^*\psi_{i,n}S_{j,i}\right] \nonumber \\
       & + & \frac{1}{\sqrt{N}}\sum_qe^{-iqn'}\left[\sum_{i,j}\sum_{n}\lambda_{j,q}^*\psi_{j,n}^*\psi_{i,n}S_{j,i}\right] \nonumber
\label{displacement}
\end{eqnarray}

Optical spectroscopy is also an important aspect of the polaron dynamics, as it can provide valuable information on various correlation functions. In this
work, the linear absorption spectra for the polaron dynamics calculated with different {\it Ans\"atze} have been studied to check
the validity of these trial wave functions. The linear absorption spectra $F(\omega)$ can be obtained by the Fourier transformation,
\begin{equation}
F(\omega) = \frac{1}{\pi}{\rm Re}\int_0^{\infty} F(t)e^{i\omega t}dt,
\label{linear_function}
\end{equation}
where $F(t)$ is the autocorrelation function of the exciton-phonon system, which is defined as
\begin{eqnarray}
F(t) &=& _{\rm ph}\langle0|_{\rm ex}\langle0| e^{i\hat{H}t}\hat{P}e^{-i\hat{H}t}\hat{P}^\dag |0\rangle_{\rm ex}|0\rangle_{\rm ph} \nonumber \\
& = & _{\rm ph}\langle0|_{\rm ex}\langle0| \hat{P}e^{-i\hat{H}t}\hat{P}^\dag |0\rangle_{\rm ex}|0\rangle_{\rm ph},
\label{auto_cor}
\end{eqnarray}
with the polarization operator
\begin{equation}
\hat{P} = \mu \sum_n(\hat{a}_n^{\dag}|0\rangle_{\rm ex}~_{\rm ex}\langle0|+|0\rangle_{\rm ex}~_{\rm ex}\langle0|\hat{a}_n^{\dag}).
\label{polarization}
\end{equation}
Details on how to calculate the linear absorption spectra of a one-dimensional exciton-phonon system from the multi-$\rm D_2$, single $\rm D_2$ and single $\rm D_1$
{\it Ans\"atze} are given in the Appendix C.

Finally, the validity of our {\it Ansatz}, Eq.~(5), will be closely scrutinized. Assuming the trial wavefunction $|{\rm D}^M_2(t)\rangle=|\Psi(t)\rangle$
at the time $t$, a deviation vector $|\delta(t)\rangle$ is then introduced to quantify the accuracy of the variational method,
\begin{equation}
|\delta(t)\rangle = \frac{\partial}{\partial t}|\Psi(t)\rangle - \frac{\partial}{\partial t}|{\rm D}^M_2(t)\rangle.
\label{deviation_1}
\end{equation}
where $|{\rm D}^M_2(t)\rangle$ and $|\Psi(t)\rangle$ obey Eq.~(\ref{motion}) and the Schr\"{o}dinger equation  $\partial |\Psi(t)\rangle / \partial t = \frac{1}{i\hbar}\hat{H}|\Psi(t)\rangle$, respectively. Using the Schr\"{o}dinger equation and the relationship $|\Psi(t)\rangle=|{\rm D}^M_2(t)\rangle$ at the moment $t$, the deviation vector $|\delta(t)\rangle$ can be calculated as
\begin{equation}
|\delta(t)\rangle = \frac{\hat{H}}{i\hbar}|{\rm D}^M_2(t)\rangle - \frac{\partial}{\partial t}|{\rm D}^M_2(t)\rangle.
\label{deviation_2}
\end{equation}
Thus, deviation from the exact Schr{\"o}dinger dynamics can be indicated by the amplitude of the deviation vector $\Delta(t)=\sqrt{\langle\delta(t)|\delta(t)\rangle}$.
In order to view the deviation in the parameter space $(W,J,g,\phi)$, a dimensionless relative deviation $\sigma$ is calculated as
\begin{equation}
\sigma = \frac{{\rm max}\{\Delta(t)\} }{{\rm mean}\{E_{\rm ph}(t)\}}, \quad \quad t \in [0, t_{\rm max}].
\label{relative_error}
\end{equation}
where the phonon energy $E_{ph}(t)$ is the main energy of the system,
and is almost the same with the amplitude of the time derivative of the wave function,
\begin{eqnarray}
N_{\rm err}(t) & = & \sqrt{-\langle\frac{\partial}{\partial t}\Psi(t)|\frac{\partial}{\partial t}\Psi(t)\rangle} \nonumber \\
& = & \sqrt{\langle {\rm D}^M_2(t)|\hat{H}^2|{\rm D}^M_2(t)\rangle} \nonumber \\
& \approx & \Delta E,
\end{eqnarray}
since $\langle E \rangle = \langle {\rm D}^M_2(t)|\hat{H}(t)|{\rm D}^M_2(t)\rangle \approx 0$ in this paper.

\begin{figure}[tbp]
\centering
\includegraphics[scale=0.4]{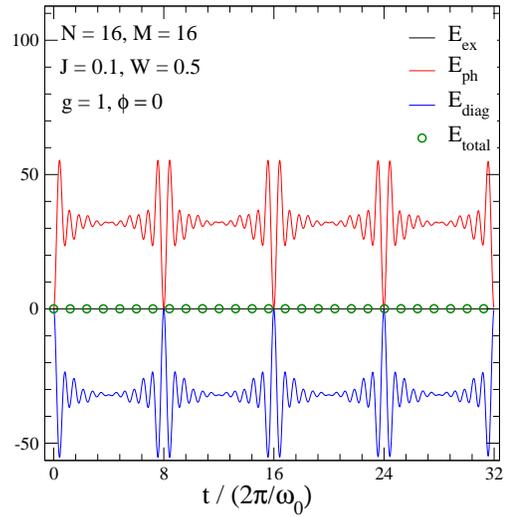}
\caption[FIG]{ $E_{\rm ex}, E_{\rm ph}, E_{\rm diag}$ and $E_{\rm total}$ obtained by the multi-${\rm D}_2$ {\it Ansatz} with $M=16$, are displayed as a function of the time $t$ for a molecular ring with $N = 16$ sites. The parameters including the transfer integral $J=0.1$, phonon energy bandwidth $W=0.5$, diagonal coupling strength $g=1$ and off-diagonal coupling strength $\phi = 0$ are set. }
\label{1ong_1}
\end{figure}

\section{Numerical results}

\subsection{Diagonal coupling}

The long-time behavior of the Holstein polaron dynamics is described by Eq.~(\ref{motion}). Fig.~\ref{1ong_1} shows the evolution of the system energies, including the total energy $E_{\rm total}$, the phonon energy $E_{\rm ph}$, the exciton energy $E_{\rm ex}$, and the exciton-phonon interaction energy $E_{\rm diag}$, for the diagonal coupling case with $J=0.1, W=0.5$, and $g=1$. For $N=16$ sites in the molecular ring, the {\it Ansatz} is formed from superposition of $M=16$  ${\rm D}_2$ wave functions, and the initial state as shown in Fig.~\ref{init}(a) is used. The periodicity of the system energies is given by $T=4\omega_0/\pi$, in perfect agreement with the expectation of $N/4W$. The finding that $E_{\rm ph} \approx -E_{\rm diag}$ and $E_{\rm ex} = E_{\rm total}\approx 0$ shows that the total energy is conserved in the Dirac-Frenkel variational dynamics.

The exciton probability $P_{\rm ex}(t, n)$ and the phonon displacement $X_{\rm ph}(t, n)$ from the multi-${\rm D}_2$ {\it Ansatz} are compared with those from the single ${\rm D}_2$ and ${\rm D}_1$ {\it Ans\"{a}tze}.
The latter can be written as
\begin{eqnarray}
&&  |{\rm D_1}(t)\rangle  =  \\ \nonumber
&& \sum_{n=1}^N \psi_{i,n}(t)\hat{a}_n^{\dag}|0\rangle_{\rm ex}\exp\left[\sum_q\left(\lambda_{n,q}\hat{b}_q^{\dag}-\lambda_{n,q}^*\hat{b}_q\right)\right] |0\rangle_{\rm ph},
\end{eqnarray}
where the displacement coefficient $\lambda_{n,q}$ is not only dependent on the moment $q$, but also on the site index $n$ in the molecular ring.
As referred in the ``Introduction," while the ${\rm D}_1$ {\it Ansatz} is effective in the diagonal coupling case,
it fails to describe the polaron dynamics of the Holstein model with off-diagonal coupling.

\begin{figure}[tbp]
\centering
\includegraphics[scale=0.4]{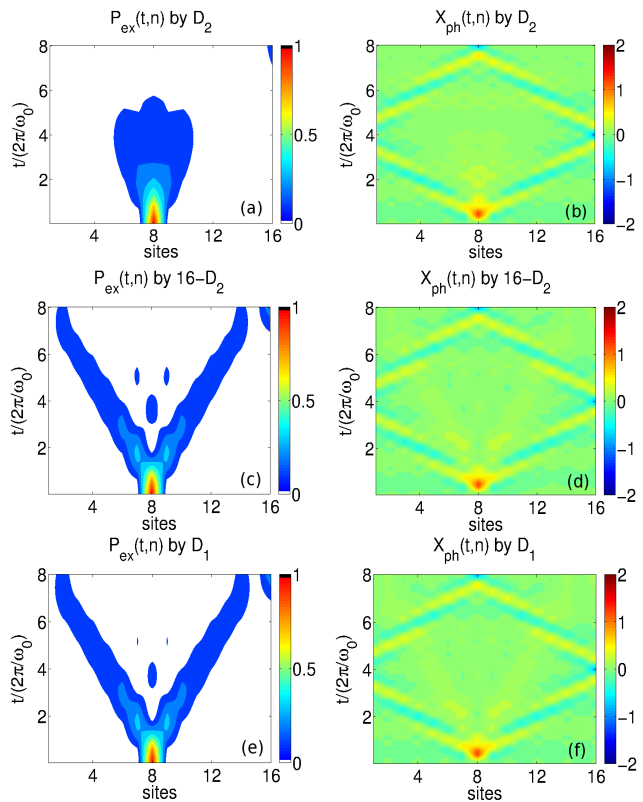}
\caption[FIG]{Time evolution of the exciton probability  $P_{\rm ex}(t, n)$ and the phonon displacement $X_{\rm ph}(t, n)$ obtained by the ${\rm D}_2$, ${\rm D}^{M=16}_2$ and ${\rm D}_1$ {\it Ans{\"a}tze} are displayed in (a)-(f) for a weak coupling case of $J=0.1, W=0.5, g = 0.1$ and $\phi = 0$.
}
\label{diagonal_1}
\end{figure}

In Fig.~\ref{diagonal_1}, the time evolution of the exciton probability $P_{\rm ex}(t, n)$ and the phonon displacement $X_{\rm ph}(t, n)$ are shown in a weak-coupling case with $g=0.1$ calculated using the ${\rm D}_2$ {\it Ansatz} [panels (a) and (b)], the ${\rm D}^{M=16}_2$ {\it Ansatz} [panels (c) and (d)], and the ${\rm D}_1$ {\it Ansatz} [panels (e) and (f)]. Similarly, time-dependent behaviors of both $P_{\rm ex}(t, n)$ and $X_{\rm ph}(t, n)$ are found to be almost the same in the multi-${\rm D_2}$ and the single ${\rm D}_1$ {\it Ans\"atze}, but at variance with those in the single ${\rm D_2}$ {\it Ansatz}. Moreover, the  propagation of the exciton wave packets can be found in Fig.~\ref{diagonal_1}(c) with the velocity $v \approx \omega_0/2\pi$, consistent with that obtained by the ${\rm D_1}$ {\it Ansatz} in Fig.~\ref{diagonal_1}(e).

\begin{figure}[tbp]
\centering
\includegraphics[scale=0.4]{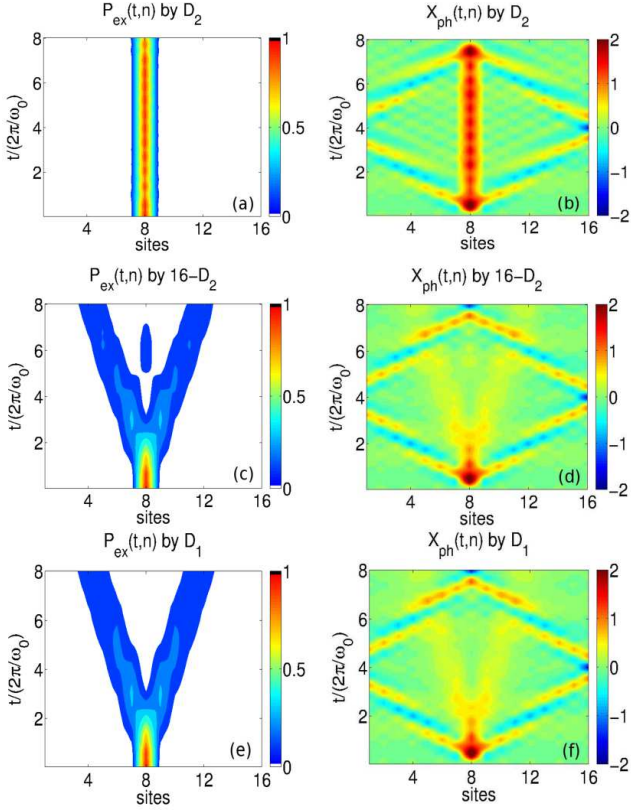}
\caption[FIG]{Time evolution of the exciton probability  $P_{\rm ex}(t, n)$ and the phonon displacement $X_{\rm ph}(t, n)$ obtained by the ${\rm D}_2$, ${\rm D}^{M=16}_2$ and ${\rm D}_1$ trial states are displayed in (a)-(f) for $J=0.1, W=0.5, g = 0.2$, and $\phi=0$.}
\label{diagonal_2}
\end{figure}

The behavior of the Holstein polaron in the intermediate diagonal coupling regime is shown in Fig.~\ref{diagonal_2} at the diagonal coupling strength $g = 0.2$. Similar time-dependent behavior in $P_{\rm ex}(t, n)$ and $X_{\rm ph}(t, n)$ is spotted in the multi-${\rm D_2}$ {\it Ansatz} and the ${\rm D}_1$ {\it Ansatz}, but not in the single ${\rm D_2}$ {\it Ansatz}. The speed of the localized exciton wave packets $v \approx \omega_0/4\pi$ is then measured from both Fig.~\ref{diagonal_2} (c) and (e) to be nearly half of that in the weak coupling case of $g=0.1$. It suggests that the velocity $v$ is inversely proportional to the diagonal coupling strength $g$. Moreover, similar phonon propagation patterns are found in (d) and (e), including the sound waves and the movement induced by the exciton-phonon interaction. The latter is found to be with the same velocity as that of the exciton wave packets.

\begin{figure}[tbp]
\centering
\includegraphics[scale=0.4]{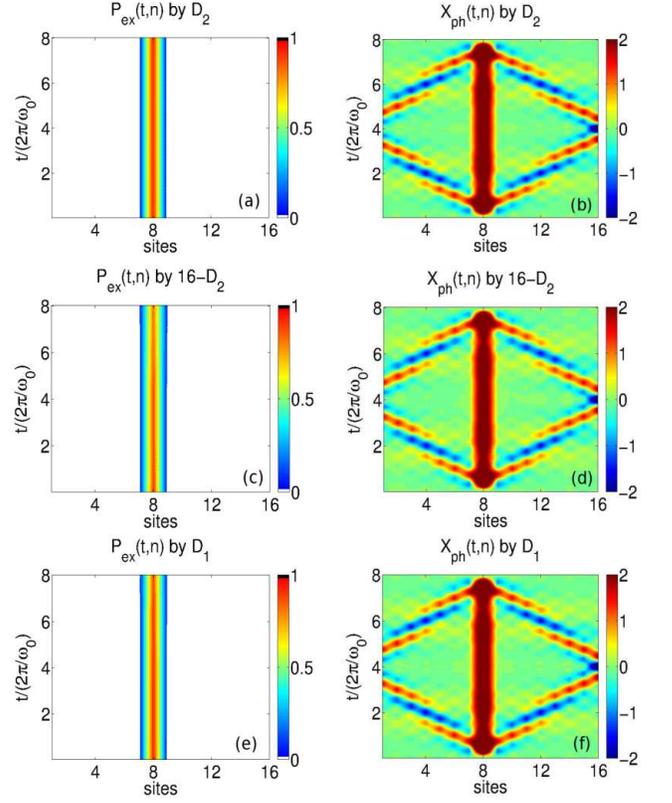}
\caption[FIG]{For the strong diagonal coupling strength $g = 0.4$, time evolution of the exciton probability  $P_{\rm ex}(t, n)$ and the phonon displacement $X_{\rm ph}(t, n)$ obtained by the ${\rm D}_2$, ${\rm D}^{M=16}_2$ and ${\rm D}_1$ {\it Ans\"atze} are displayed in (a)-(f) at $J=0.1, W=0.5$ and $\phi=0$.
}
\vspace{1.5\baselineskip}
\label{diagonal_3}
\end{figure}

The results of the variational dynamics in the strong coupling case with $g=0.4$ are shown in Figs.~\ref{diagonal_3}(a)-(f), corresponding to the  ${\rm D}_2$, ${\rm D}^{M=16}_2$ and ${\rm D}_1$ {\it Ans\"atze}, respectively. Different from the weak and intermediate coupling cases in Figs.~\ref{diagonal_1} and ~\ref{diagonal_2}, all of the exciton probabilities $P_{\rm ex}(t, n)$ and phonon displacements $X_{\rm ph}(t, n)$ obtained by these three kinds of the trial wave functions nearly identical, indicating that all of the ${\rm D}_2$, multi-${\rm D}_2$ and ${\rm D}_1$ {\it Ans\"atze} are effective in the strong diagonal coupling case.


\begin{figure}[tbp]
\centering
\includegraphics[scale=0.4]{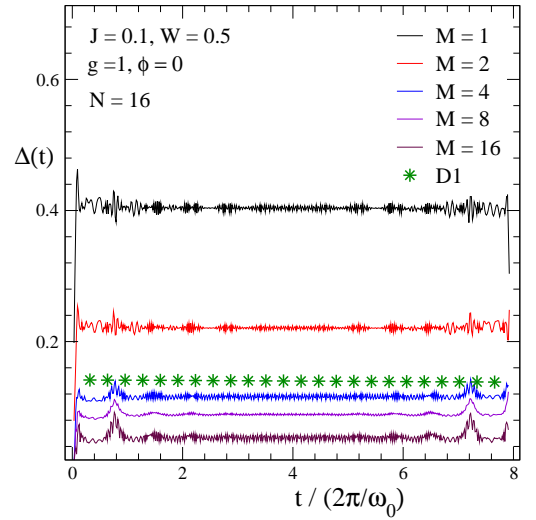}
\caption[FIG]{The amplitude of the deviation vector $\Delta(t)$ from the Schr\"{o}dinger equation is plotted as a function
of the time $t$ with the unit $2\pi/\omega_0$ at $J=0.1, W=0.5, g=1.0$ and $\phi = 0$.
Lines represent the results obtained from the multi-${\rm D_2}$ {\it Ansatz} with various values of $M$, and the stars denote those from the $\rm {D}_1$ {\it Ansatz}.
}
\label{error_1}
\end{figure}

Using the amplitude of the deviation vector $\Delta(t)$ from the exact Schr{\"o}dinger dynamics, the validity of the multi-${\rm D_2}$ {\it Ansatz} is
investigated for various numbers of $M$ defined in Eq.~(\ref{wave}), as shown in Fig.~\ref{error_1}.  The amplitude of $\Delta(t)$ is almost constant, and the
time-averaged value $\langle\Delta(t)\rangle$ monotonically decreases with $M$. It indicates that the multi-${\rm D_2}$ trial state approaches the exact solution to the Schr{\"o}dinger equation when $M$ is increased. For comparison, $\Delta(t)$ obtained by the ${\rm D_1}$ {\it Ansatz} is also plotted. The time-average value of $\Delta(t)$ is smaller than that obtained from the single ${\rm D_2}$ {\it Ansatz} ($M=1$), consistent with our previous contention that ${\rm D_1}$ {\it Ansatz} is more accurate than ${\rm D_2}$  {\it Ansatz} in the diagonal coupling regime. Interestingly, $\langle\Delta(t)\rangle$ calculated from the multi-${\rm D_2}$ {\it Ansatz} with $M=16$ is $0.05$, much smaller than $0.14$ from the ${\rm D_1}$ {\it Ansatz} and $0.40$ from the single ${\rm D_1}$ {\it Ansatz}, showing the superiority of the multi-${\rm D_2}$ {\it Ansatz}.

\begin{figure}[bp]
\centering
\includegraphics[scale=0.4]{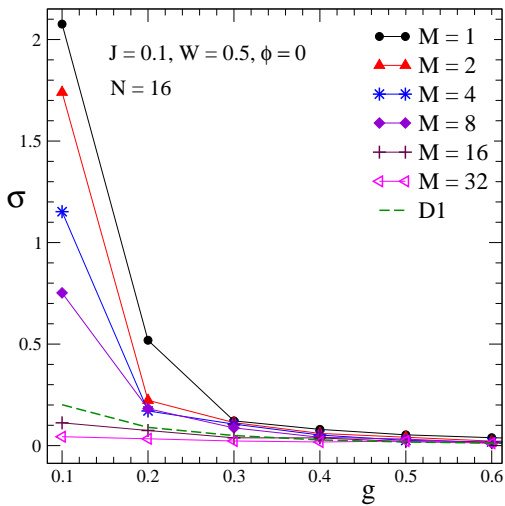}
\caption[FIG]{The relative deviation $\sigma$ defined in Eq.~(\ref{relative_error}) is displayed as a function
of the diagonal coupling strength $g$ at $J=0.1, W=0.5$ and $\phi = 0$. The lines with dots represent the results obtained from the multi-${\rm D_2}$ {\it Ansatz} with various values of $M$, and the dashed line denotes those from the $\rm {D}_1$ {\it Ansatz}. The size of the molecular ring is $N=16$.
}
\label{error_2}
\end{figure}

Among the three trial states, the validity test of the multi-${\rm D_2}$ states is comprehensively performed for various values of the diagonal coupling strength $g$ and transfer integral $J$ in the diagonal coupling only regime. In Fig.~\ref{error_2}, the relative deviation $\sigma$ defined in Eq.~(\ref{relative_error}) is displayed as a function of the  diagonal coupling strength $g$ for various values of $M$ for the case of $J=0.1, W=0.5$ and $\phi = 0$. As $M$ increases, the relative error $\sigma$ decreases, especially for the weak coupling case of $g=0.1$. For comparison, $\sigma$ calculated from the ${\rm D_1}$ {\it Ansatz} is also shown, with $\sigma$  obviously larger than that of the multi-${\rm D_2}$ {\it Ansatz} with $M=32$, further supporting the superiority of the multi-${\rm D_2}$ {\it Ansatz} over the ${\rm D_1}$ state. Moreover, $\sigma \approx 0.01$ at $M=32$ indicates that the variational method based on the multi-$\rm D_2$ {\it Ansatz} is possible to be numerically exact in the limit of $M \rightarrow \infty$, where ``numerically exact" means the relative error $\sigma=0$ within numerical errors.

\begin{figure}[tbp]
\centering
\includegraphics[scale=0.4]{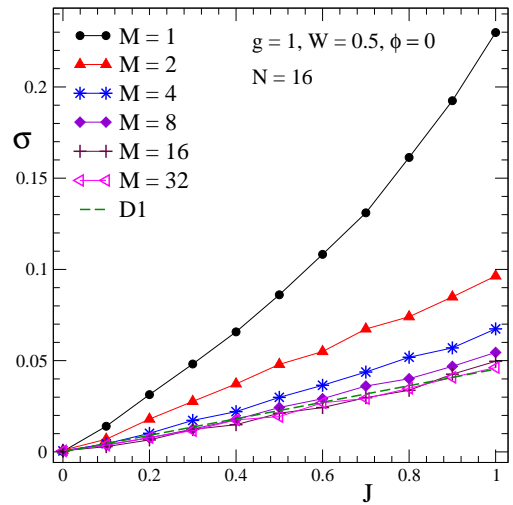}
\caption[FIG]{The relative deviation $\sigma$ in a $16$-site molecular ring is displayed as a function
of the transfer integral $J$ in the case of $g=1, W=0.5$ and $\phi = 0$.
The lines with dots represent the results obtained from the multi-${\rm D_2}$ {\it Ansatz} with various values of $M$, and the dashed line denotes those from the $\rm {D}_1$ {\it Ansatz}.
}
\label{error_3}
\end{figure}

The behavior of the relative error $\sigma$ as a function of the transfer integral $J$ is also investigated for several values of $M$. As shown in Fig.~\ref{error_3}, $\sigma$
monotonically increases with $J$. When $J$ is larger, the wavefunction approaches a plane wave, which is difficult to be described by superpositions of coherent states. With an increase in $M$, however, the relative deviation $\sigma$ is monotonically reduced indicating the improved efficiency of the ${\rm D_2}$
{\it Ansatz} even in the case with a large transfer integral. Moreover, the results show that the multi-${\rm D_2}$ {\it Ansatz} is at least as accurate as the ${\rm D_1}$ {\it Ansatz} if $M=32$ superpositions are used.

\subsection{Off-diagonal coupling}

\begin{figure}[tbp]
\centering
\includegraphics[scale=0.4]{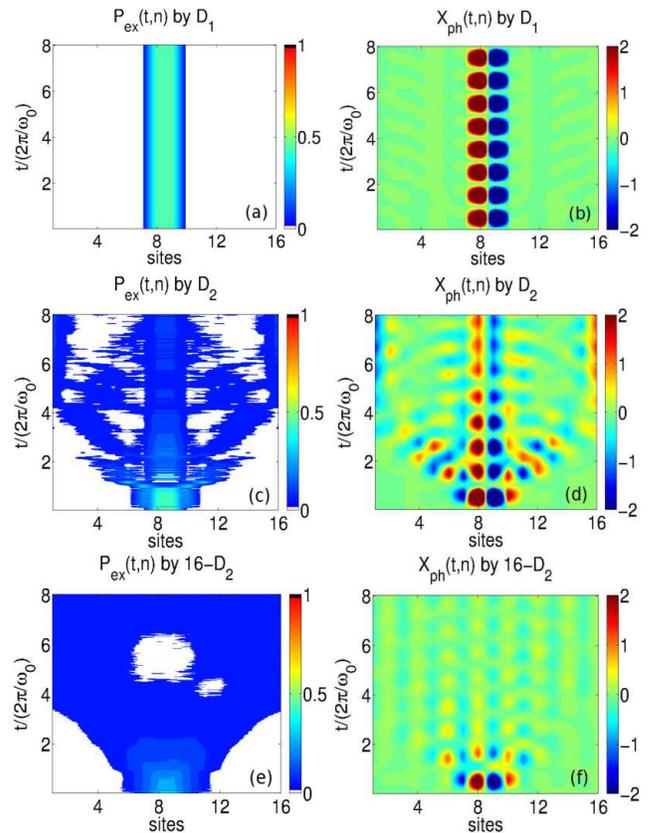}
\caption[FIG]{In the off-diagonal coupling case with $\phi = 1$ and $J=W=g=0$, time evolution of the exciton probability  $P_{\rm ex}(t, n)$ and phonon displacement $X_{\rm ph}(t, n)$ obtained by the $1{\rm D}_1$, ${\rm D}_2$ and ${\rm D}^{M=16}_2$ {\it Ans{\"a}tze} are displayed in (a)-(f).
}
\label{off-diagonal_1}
\end{figure}

In this section, we probe the dynamics of the Holstein polaron in the off-diagonal coupling regime
via the multi-${\rm D_2}$ {\it Ansatz}, in comparison with those obtained by the single ${\rm D_2}$ and the single ${\rm D_1}$ {\it Ans\"{a}tze}. The initial state takes the double excited states shown in Fig.~\ref{init}(b). For simplicity, only the off-diagonal coupling is considered in the simulations. Time evolution of the exciton probability $P_{\rm ex}(t, n)$ and phonon displacement $X_{\rm ph}(t, n)$ obtained by the single ${\rm D}_1$, the single ${\rm D}_2$ and ${\rm D}_2^{M=16}$ {\it Ans\"atze} are displayed. In Fig.~\ref{off-diagonal_1}(a) and (b) for the ${\rm D_1}$ {\it Ansatz}, one can find that self-trapping occurs at the $8$-th and $9$-th sites, at variance with the delocalization expectation of the Holstein dynamics in the off-diagonal coupling case. It is consistent with the previous impression that ${\rm D_1}$ {\it Ansatz} fails to describe the dynamics of the Holstein polaron in the off-diagonal coupling regime.

In contrast, the spread of the exciton is found in the results obtained by the single ${\rm D_2}$ and multi-$\rm D_2${\it Ans\"atze}. However, as shown in Figs.~\ref{off-diagonal_1}(c) and (d), the wave function obtained by the single ${\rm D_2}$ {\it Ansatz} is still localized, different from  those obtained by the multi-$D^{M=16}_2$ {\it Ansatz} shown in the Figs.~\ref{off-diagonal_1}(e) and (f). It indicates that the multi-${\rm D_2}$ {\it Ansatz} is more effective than the single ${\rm D_2}$ {\it Ansatz} for depicting the Holstein polaron dynamics in the off-diagonal coupling case. Moreover, one can find three different stages of the exciton motion in figure~\ref{off-diagonal_1}(e), separated by the characteristic time
$t_1 \approx 2\pi/\omega_0$ and $t_2 \approx 8\pi/\omega_0$. When $t < t_1$, the exciton is self-trapped in the initial state, pointing to the localized exciton state.
After that, the exciton starts to spread over the molecular ring with the velocity $v \approx \omega_0/\pi$. Until $t \geq t_2$, a uniform distribution of the
exciton wave packets appears, indicating that the exciton is in a delocalized state.

\begin{figure}[tbp]
\centering
\includegraphics[scale=0.4]{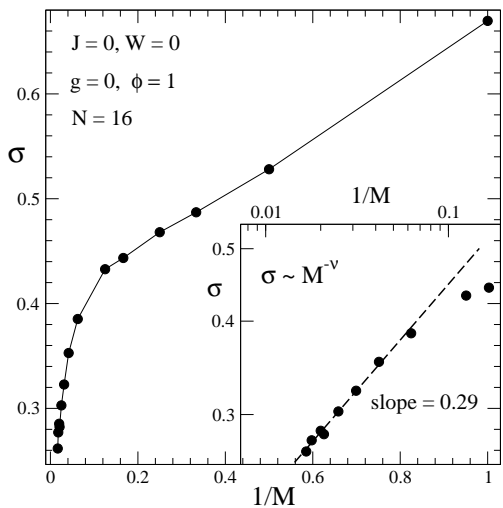}
\caption[FIG]{The relative deviation $\sigma$ from the multi-${\rm D}_2$ {\it Ansatz} is displayed as a function of $1/M$ for the the off-diagonal coupling case with
the strength $\phi=1$, and other parameters a$J=W=g=0$ are set. The size of the molecular ring is $N=16$. In the inset, the relationship $\sigma \sim M^{-\nu}$ is displayed
on a log-log scale, and the dashed line represents a power-law fit.
}
\label{off-diagonal_2}
\vspace{2.5\baselineskip}
\end{figure}

Via the relative deviation $\sigma$ defined in Eq.~(\ref{relative_error}), the validity of the multi-${\rm D_2}$ {\it Ansatz} can be further confirmed.
In Fig.~\ref{off-diagonal_2}, the relative deviation $\sigma$ is plotted as a function of $1/M$ for the off-diagonal coupling only case with $\phi=1.0$.
Both the transfer integral $J$ and the phonon bandwidth are set to be zero. As $M$ increases, the relative deviation $\sigma$
decreases and approaches zero as $M$ goes to infinity. For example, the value of $\sigma(M=60)=0.26$ is much smaller than $\sigma(M=1)=0.67$.
According to the fitting in the inset, the relationship $\sigma \sim M^{-\nu}$ is revealed with
the exponent $\nu = 0.29(1)$, further confirming the prediction $\sigma = 0$ in the limit of $M \rightarrow \infty$.
Hence, it can be concluded that the variational method based on the multi-${\rm D_2}$ {\it Ansatz} is possible to be numerically exact ($\sigma=0$) in both of the diagonal and off-diagonal coupling regimes. Since any quantum state of a system of multiple oscillators (boson modes) can be represented by a continuous superposition of coherent states (often referred to as the unit decomposition property of the aforementioned states), there is a good chance that the multi-$\rm D_2$ {\it Ansatz} is numerically exact, i.e., provides exact results in the $M \rightarrow \infty$  limit. However, it remains unclear how practical is the above statement, since the values of the multiplicity $M$, needed for the {\it Ansatz} to converge to the exact solution of the dynamical Schr\"{o}dinger equation might be unrealistically large. The above question will be addressed elsewhere.

\begin{figure}[tbp]
\centering
\includegraphics[scale=0.4]{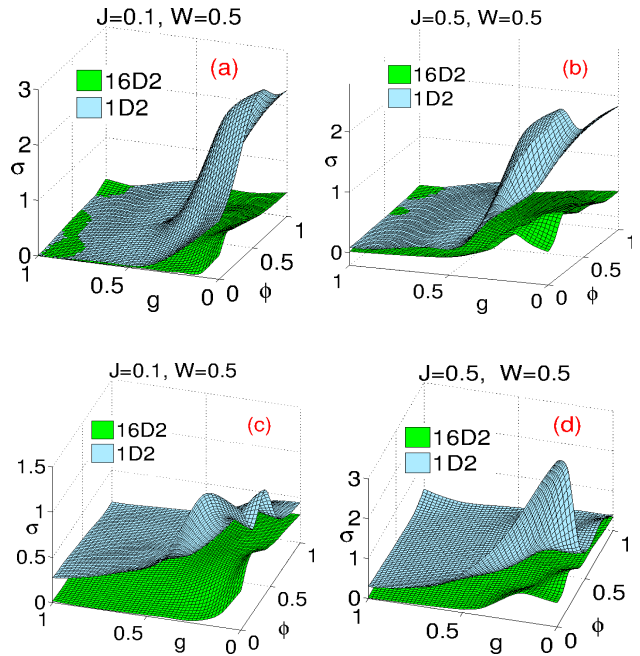}
\caption[FIG]{ The relative deviation $\sigma$ is displayed for the single ${\rm D_2}$ and multi-${\rm D_2}$ {\it Ans\"atze} as a function of the diagonal coupling
strength $g$ and off-diagonal coupling strength $\phi$. In (a) and (b), the exciton at $t=0$ is created on a single site $n=N/2$, while in (c) and (d),
a two-site occupied state is used as the initial state.
}
\label{global}
\end{figure}

Finally, accuracy of the multi-${\rm D_2}$ {\it Ansatz} is quantified for the parameter regime $0\leq g,\phi \leq 1$, in comparison with that of the single-${\rm D_2}$ {\it Ansatz}, as shown in Fig.~\ref{global}. The influence of the excitonic initial state is also taken into account. Figs.~\ref{global} (a) and (b) correspond to the one-site occupied initial
state shown in Fig.~\ref{init}(a), and Figs.~\ref{global} (c) and (d) to the two-site occupied initial states shown in Fig.~\ref{init}(b). For each initial state, two different values of the transfer integral $J=0.1$ and $0.5$ are used. Our results show that $D^{M=16}_2$ {\it Ansatz} deviates little from the exact solutions of the time-dependent Schr\"odinger equation in the whole parameter regime for both the two initial states and the cases with small and large transfer integral, thereby further confirming the validity of the variational method. Moreover, the significant improvement of the validity for the multi-${\rm D_2}$ {\it Ansatz} from the single ${\rm D_2}$ {\it Ansatz} is found especially in the weak diagonal or off-diagonal coupling regimes, confirming the high accuracy of the {\it Ansatz}.

\begin{figure}[tbp]
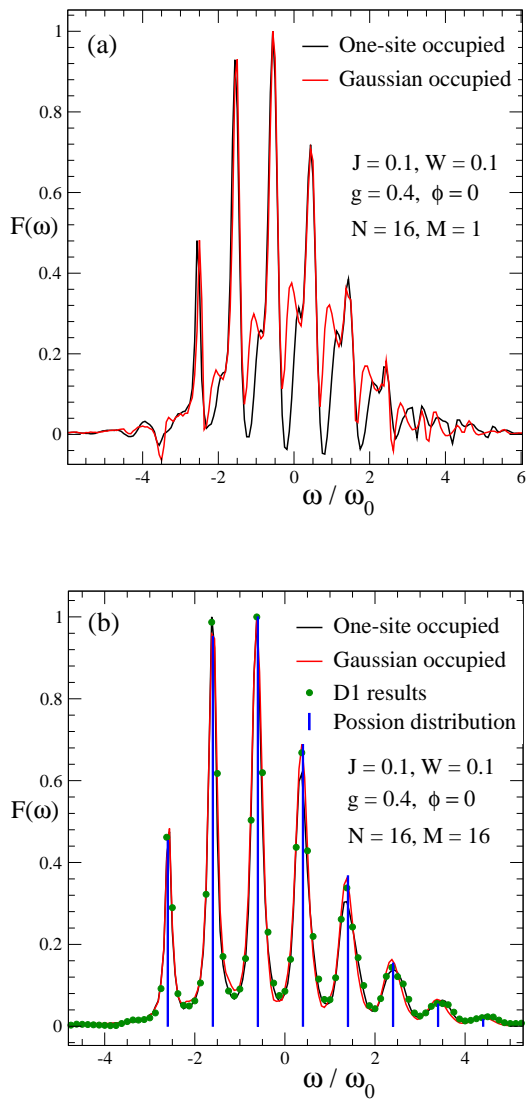

\centering
\includegraphics[scale=0.4]{linear_a2.eps} \\
\vspace{2.5\baselineskip}
\includegraphics*[scale=0.4]{linear_b2.eps}
\caption[FIG]{ Linear absorption spectra of a $16$-sites, one-dimensional ring of a coupled exciton-phonon system are displayed in (a)
for the single $D_2$ {\it Ansatz}, and in (b) for the $D_2^{M=16}$ {\it Ansatz}. Two kinds of the initial states including
the one-site occupied and Gaussian occupied exciton distributions are used in the simulations, and the Huang-Rhys factor is $S=2.56$. For comparison,
the numerical results obtained by the single $\rm D_1$ {\it Ansatz} and the fitting of a Poisson distribution are given in (b) with the solid circles and
bars, respectively. A rescaled factor is applied to normalize the spectral maxima for facilitate comparisons.
}
\label{absorption spectra}
\end{figure}

\subsection{Absorption spectra}

Besides the relative deviation, the validity of the {\it Ans\"atze} in providing reliable dynamical information can also be gauged by the accuracy of optical spectra,
as analytical expressions for the absorption and fluorescence spectra are well-known if the transfer integral $J$ and the phonon bandwidth $W$ are negligible. In this study,
the set of the parameters $J=0.1, W=0.1, g=0.4$ and $\phi = 0$ is used, and the Huang-Rhys factor $S$ is then calculated according to the relaxation energy defined by
\begin{equation}
E_r \equiv \int_{-\infty}^{\infty} {\frac{C_{00}\left(\omega\right)}{\omega}}\,d\omega=\sum_{q}g_q^2\omega_{q}\equiv S\omega_{0},
\label{huang_rhys}
\end{equation}
where $\omega_{0}$ is the central energy of the phonon band, $g_q=g=0.4$ is the diagonal coupling, and $\omega_q$ is the frequency at the moment $q$. From this equation, we can obtain $S=2.56$ corresponding to the diagonal coupling strength $g=0.4$. Two types of initial states, one-site-occupied excited state and the excitation with a Gaussian-type distribution spanned on $7$ sites, have been applied in the study of optical spectra. To facilitate comparisons, spectral maxima are normalized to unity.

Three trial states including the single $\rm D_1$, the single $\rm D_2$ and the multi-$\rm D_2$ {\it Ans\"atze} are investigated, as they differ in terms of the variational parameters in describing the phonon behavior. Linear absorption spectrum is a very useful indicator of the {\it Ansatz} validity in the investigation of the dynamics of a polaron system. For the single $\rm D_2$ {\it Ansatz}, the phonon displacement is only described by one set of variational parameters, leading to the lack of exciton-phonon correlation between exciton and phonon, and eventually to the absence of long-range correlation in the autocorrelation function and an inaccurate description of optical spectra. Only for the cases with the one-site occupied initial state under strong diagonal coupling and small $J$ where the exciton is localized, as shown by the black solid line in Fig.~\ref{absorption spectra}(a), the inability of the single $\rm D_2$ {\it Ansatz} is alleviated. As for the red line, where the initial electronic excitation adopts a Gaussian distribution, the spectrum even exhibits negative values around $\omega=-3.5\omega_0$. In contrast, the single $\rm D_1$ and the multi-$\rm D_2$ {\it Ans\"atze} guarantee the long-range exciton-phonon correlation, therefore can provide accurate absorption spectra. Fig.~\ref{absorption spectra}(b) shows similar correct spectra for both single D$_1$ and multi-D$_2$ {\it Ans\"atze}.

According to the Huang-Rhys theory, the phonon sidesbands at zero temperature follow a Poisson distribution,
\begin{equation}
F \left(\omega\right)=\exp\left(-S\right)\sum_{n=0}^{\infty}{\frac{{S^{n}}}{n!}}\delta\left(\omega+E_r-n\omega\right)
\label{Possion}
\end{equation}
From Eq.~(\ref{Possion}), the left most sideband, $n=0$, corresponding to the zero-phonon line, should be located at $\omega=-S\omega_{0}$, ie. $\omega=-2.56\omega_0$
consistent with our result $\omega=-2.64\omega_0$ as shown in Fig.~\ref{absorption spectra}(b). What is more, the tallest peaks at $n=1$ and $2$ show almost same height, in agreement with the predict that tallest of phonon sidebands should be $n=S-1=1.56$ peak when $S\gg1$. Further, using a fitting method, the bar plot of the Poisson distribution with the parameter $\lambda=2.2$ is given, consistent with our spectra obtained from time-dependent evolution of single D$_1$ and multi-D$_2$ {\it Ans\"atze}.
The slight deviation of the Poisson parameter $\lambda$ from the Huang-Rhys factor $S=2.56$ is due to the nonzero values of $W$ and $J$.

\section{Conclusion}

Using the multi-$\rm D_2$ {\it Ansatz} as the trial state, we have systematically studied the dynamics of a one-dimensional Holstein polaron with simultaneous diagonal and off-diagonal exciton-phonon coupling via the Dirac-Frenkel time-dependent variational approach. Compared to the single $\rm D_2$ {\it Ansatz}, the multi-$\rm D_2$ counterpart is much more sophisticated and contains more flexible variational parameters, leading to superior quality simulations of polaron dynamics. Special attention is paid to testing the validity of our time-dependent variational approach by quantifying how closely the trial state follows the Schr\"odinger equation. Linear absorption spectra derived from the trial state are also studied as a sensitive indicator of the {\it Ansatz} validity in the investigation of polaron dynamics.

Our numerical results show considerable improvements in accuracy of polaron dynamics by the multi-$\rm D_2$ {\it Ansatz}, in comparison with the usual, single D$_1$ and D$_2$ trial states. In the diagonal coupling regime, the multi-$\rm D_2$ {\it Ansatz} is found to be at least as accurate as the single $\rm D_1$ {\it Ansatz} for various values of the diagonal-coupling strength $g$ and the transfer integral $J$, and remarkably better than the single D$_2$ {\it Ansatz}. In the off-diagonal coupling regime, however, the multi-$\rm D_2$ {\it Ansatz} is shown to be much more potent in depicting the Holstein polaron dynamics than the single D$_2$ {\it Ansatz}, while the single D$_1$ {\it Ansatz} fails completely. As the number of superposed states $M$ increases, one can find visible decays of the relative
deviation $\sigma$ in the weak diagonal coupling regime as well as the off-diagonal coupling regime, confirming respectable accuracies of the multi-D$_2$ {\it Ansatze}. Moreover, $\sigma=0$ is predicted by the numerical fitting in the limit of $M \rightarrow \infty$, inferring that the Dirac-Frenkel time-dependent variational approach based on the multi-$\rm D_2$ is possible to be numerically exact.

The single Davydov ${\rm D_1}$ {\it Ansatz} is a trial state with sufficient flexibilities to handle accurately quantum dynamics from the celebrated spin-boson model to large light-harvesting complexes in photosynthesis \cite{wu_duan,Ye_12}. Very recently, a systematic coherent-state expansion of the ground state wave function that is based on the Davydov ${\rm D_1}$ {\it Ansatz}, which we shall call the ``multi-${\rm D_1}$ {\it Ansatz}," is developed for a number of models \cite{zhou_14, bera_14, bera_14a}.
It is a generalization of a variational wave function originally proposed by Silbey and Harris \cite{sil84}, and also an extension of the hierarchy
of translationlly-invariant {\it Ans\"{a}tze} proposed by Zhao {\it et al.}~\cite{zhao_97,zhao_97b}. The results of the quantum phase transition
obtained from the multi-${\rm D_1}$ {\it Ansatz} are more accurate than that from the single ${\rm D_1}$ {\it Ansatz},
and they are in agreement with DMRG and ED results, showing the superiority of the multi-${\rm D_1}$ {\it Ansatz}.
The successful application of the Multi-D$_2$ {\it Ansatz} in the Holstein polaron dynamics points to the possibility that the multi-${\rm D_1}$ {\it Ansatz} is
not only valid for studying static properties of the model Hamiltonians, but also holds promise to their dynamics simulation. Our work here therefore serves as a proof of concept.
Dynamics simulation of a Holstein polaron by the multi-$\rm D_2$ {\it Ansatz} has convincingly shown that even a relatively simple wave function such as the Davydov D$_2$ {\it Ansatz}, when used in an expandable superposition,
can still produce superior results. If the D$_2$ {\it Ansatz} were to be replaced by the more sophisticated $\rm D_1$ trial state, and applied in a multitude as demonstrated here, much better results can be expected on simulating quantum dynamics of many-body systems. Work in this direction is now in progress.

\section*{Acknowledgments}

Support from the Singapore National Research Foundation through the Competitive Research Programme (CRP) under Project No.~NRF-CRP5-2009-04
is gratefully acknowledged. One of us (NJZ) is also supported in part by National Natural Science Foundation of China under Grant No.~$11205043$.

\appendix
\section{Time evolution of the multi-${\rm D}_2$ trial state}
As mentioned in Sec.II ``Methodology", the time evolution for the multi-$\rm D_2$ {\it Ansatz} can be derived
by employing Dirac-Frenkel time-dependent variational method. According to the definition of the multi-$\rm D_2$ $\it Ansatz$ in Eq.~(\ref{wave}) and
the Dirac-Frenkel variational principle, the variational parameters $\psi_{i,n}\left(t\right)$ and $\lambda_{i,q}\left(t\right)$ should obey
\begin{eqnarray}
\frac{d}{dt}\left(\frac{\partial L}{\partial\dot{\psi_{i,n}^{\ast}}}\right)-\frac{\partial L}{\partial\psi_{i,n}^{\ast}} & = & 0, \nonumber \\
\frac{d}{dt}\left(\frac{\partial L}{\partial\dot{\lambda_{k,q}^{\ast}}}\right)-\frac{\partial L}{\partial\lambda_{k,q}^{\ast}} & = & 0,
\label{A1}
\end{eqnarray}
where $L$ is the Lagrangian defined in Eq.~(\ref{Lagrangian}). Inside the Lagrangian, the first term reads as following,
\begin{eqnarray}
&& \left\langle {\rm D}^M_2\left(t\right)\right|\frac{\overrightarrow{\partial}}{\partial t}\left|{\rm D}^M_2\left(t\right)\right\rangle -\left\langle {\rm D}^M_2\left(t\right)\right|\frac{\overleftarrow{\partial}}{\partial t}\left|{\rm D}^M_2\left(t\right)\right\rangle \nonumber \\
&& =  \sum_{i,j}^{M}\sum_{n}\left(\psi_{jn}^{\ast}\dot{\psi}_{in}-\dot{\psi}_{jn}^{\ast}\psi_{in}\right)S_{ji} \nonumber \\
&& +\sum_{i,j}^{M}\sum_{n}\psi_{jn}^{\ast}\psi_{in}\sum_{q}\left[\frac{1}{2}(\dot{\lambda}_{jq}^{\ast}\lambda_{jq}+\lambda_{jq}^{\ast}\dot{\lambda}_{jq}) \right. \nonumber \\
 &&\left.-\frac{1}{2}(\dot{\lambda}_{iq}\lambda_{iq}^{\ast}+\lambda_{iq}\dot{\lambda}_{iq}^{\ast})+\lambda_{jq}^{\ast}\dot{\lambda}_{iq}-\lambda_{iq}\dot{\lambda}_{jq}^{\ast}\right]S_{ji},
\label{A2}
\end{eqnarray}
and the second term is
\begin{eqnarray}
\langle D^M_2(t)|\hat{H}|D^M_2(t)\rangle &=& E_{\rm ex} +E_{\rm ph}+E_{\rm diag}+E_{\rm off},
\label{A3}
\end{eqnarray}
where $E_{\rm ex}, E_{\rm ph},E_{\rm diag}$ and $E_{\rm off}$ denote the energies of the exciton, phonon, diagonal couple term and off-diagonal coupling term, respectively,
which can be calculated based on Eq.~(\ref{energy}).

Time partial derivatives of $\lambda_{iq}(t)$ and $\psi_{in}(t)$
are then obtained by substituting Eqs.~(\ref{A2}) and (\ref{A3}) into Eq.~(\ref{A1}), which are
\begin{eqnarray}
\label{xxx}
&& -i\sum_{i}^{M}\dot{\psi}_{in}S_{ki}  \nonumber \\
&& -\frac{i}{2}\sum_{i}^{M}\psi_{in}\sum_{q}\left(2\lambda_{kq}^{\ast}\dot{\lambda}_{iq}-\dot{\lambda}_{iq}\lambda_{iq}^{\ast}-\lambda_{iq}\dot{\lambda}_{iq}^{\ast}\right)S_{k,i}\nonumber\\
&=&~J\sum_{i}^{M}\left(\psi_{i,n+1}+\psi_{i,n-1}\right)S_{ki}  \\
&& -\sum_{i}^{M}\psi_{in}\sum_{q}\omega_{q}\lambda_{kq}^{\ast}\lambda_{iq}S_{ki} \nonumber \\
&& +g\sum_{i}^{M}\psi_{in}\sum_{q}\omega_{q}\left(e^{iqn}\lambda_{iq}+e^{-iqn}\lambda_{kq}^{\ast}\right)S_{ki} \nonumber \\
&& -\frac{1}{2}\phi\sum_{i}^{M}\sum_{q}\omega_{q}\psi_{i,n+1}\left[e^{iqn}(e^{iq}-1)\lambda_{iq} \nonumber \right. \\
&&\left.+e^{-iqn}(e^{-iq}-1)\lambda_{kq}^{\ast}\right]S_{ki} \nonumber \\
&& -\frac{1}{2}\phi\sum_{i}^{M}\sum_{q}\omega_{q}\psi_{i,n-1}\left[e^{iqn}(1-e^{-iq})\lambda_{iq} \right. \nonumber \\
&&\left.+e^{-iqn}(1-e^{iq})\lambda_{kq}^{\ast}\right]S_{ki}, \nonumber
\end{eqnarray}
and
\begin{eqnarray}
\label{A5}
&& -i\sum_{i}^{M}\sum_{n}\psi_{kn}^{\ast}\dot{\psi}_{in}\lambda_{iq}S_{ki} \nonumber \\
&& -i\sum_{i}^M\sum_{n}\psi_{kn}^{\ast}\psi_{in}\dot{\lambda}_{iq}S_{ki} - \frac{i}{2}\sum_{i}^{M}\sum_{n}\psi_{kn}^{\ast}\psi_{in}\lambda_{iq}S_{k,i} \nonumber \\
&&\sum_{p}\left(2\lambda_{kp}^{\ast}\dot{\lambda}_{ip}-\dot{\lambda}_{ip}\lambda_{ip}^{\ast}-\lambda_{ip}\dot{\lambda}_{ip}^{\ast}\right) \nonumber \\
& = & ~J\sum_{i}^{M}\sum_{n}\psi_{kn}^{\ast}\left(\psi_{i,n+1}+\psi_{i,n-1}\right)\lambda_{iq}S_{k,i}  \\
&& -\sum_{i}^{M}\sum_{n}\psi_{kn}^{\ast}\psi_{in}\left(\omega_{q}+\sum_{p}\omega_{p}\lambda_{kp}^{\ast}\lambda_{ip}\right)\lambda_{iq}S_{ki} \nonumber \\
&& +g\sum_{i}^{M}\sum_{n}\psi_{kn}^{\ast}\psi_{in}\omega_{q}e^{-iqn}S_{ki} \nonumber \\
&& +g\sum_{i}^{M}\sum_{n}\psi_{kn}^{\ast}\psi_{in}\lambda_{iq}\sum_{p}\omega_{p}\left(e^{ipn}\lambda_{ip}+e^{-ipn}\lambda_{kp}^{\ast}\right)S_{k,i} \nonumber \\
&& -\frac{1}{2}\phi\sum_{n}\sum_{i}^{M}\omega_{q}\psi_{kn}^{\ast}\left[\psi_{i,n+1}e^{-iqn}(e^{-iq}-1) \right. \nonumber \\
&&\left.+\psi_{i,n-1}e^{-iqn}(1-e^{iq})\right]S_{ki} \nonumber \\
&& -\frac{1}{2}\phi\sum_{n}\sum_{i}^{M}\left(\psi_{k,n+1}^{\ast}\psi_{i,n}+\psi_{kn}^{\ast}\psi_{i,n+1}\right)\lambda_{iq} \nonumber\\
&&\sum_{p}\omega_{p}\left[e^{ipn}(e^{ip}-1)\lambda_{ip}+e^{-ipn}(e^{-ip}-1)\lambda_{kp}^{\ast}\right]S_{k,i}, \nonumber
\end{eqnarray}
where $S_{k,i}$ is the Debye-Waller factor defined in Eq.~(\ref{factor}). Unfortunately, both Eqs.~(\ref{xxx}) and (\ref{A5}) contain
the coupling time partial derivatives of $\lambda_{iq}(t)$ and $\psi_{in}(t)$ with the form $\sum_i A_1\dot{\psi}_{in} + \sum_{i,q}A_2\dot{\lambda}_{iq} +
\sum_{i,q}A_3\dot{\lambda}_{iq}^{\ast}=B$, where $A_1, A_2, A_3$ and $B$ are coefficient vectors, irrelative to the time partial derivatives of the variational parameters.
By numerically solving these linear equations at each time $t$, one can calculate the values of $\dot{\psi}_{in}$ and $\dot{\lambda}_{iq}$ accurately.
Runge-Kutta $4$-th order method is then adopted for the time evolution of the Holstein polaron, including the energies
$E_{\rm total}(t)=E_{\rm ex} +E_{\rm ph}+E_{\rm diag}+E_{\rm off}$, the normalization $Norm(t)$, the exciton probability $P_{\rm ex}(t, n)$ and
phonon displacement $X_{\rm ph}(t, n)$.

Finally, the amplitude of the deviation vector $\Delta(t)$ defined in Eq.~(\ref{deviation_1}) is calculated as
\begin{eqnarray}
\label{deviation_3}
\Delta^2(t) &=& \left\langle \delta \left(t\right)\vert\delta \left(t\right)\right\rangle  \nonumber \\
 & = & \sum_{n}\sum_{i,j}^{M}\left[\sum_{q}\left(A_{jnq}^{*}+B_{jnq}^{*}\lambda_{iq}\right)\right]S_{ji}  \\
 &  &\left[\sum_{q}\left(A_{inq}+B_{inq}\lambda_{jq}^{\ast}\right)\right]+\sum_{n,q}\sum_{i,j}^{M}B_{jnq}^{*}S_{ji}B_{inq}, \nonumber
\end{eqnarray}
where the matrixes $A_{jnq}$ and $B_{jnq}$ are respectively defined as

\begin{eqnarray*}
 &   & A_{jnq} \\
 & = & ~i\frac{1}{N}\dot{\psi}_{j,n}\left(t\right)\\
 &  & -i\frac{1}{2}\psi_{j,n}\left(t\right)\left[\dot{\lambda}_{j,q}\left(t\right)\lambda_{j,q}^{*}\left(t\right)+\lambda_{j,q}\left(t\right)\dot{\lambda}_{j,q}^{*}\left(t\right)\right]\\
 &  & +J\frac{1}{N}\left[\psi_{j,n+1}\left(t\right)+\psi_{j,n-1}\left(t\right)\right]\\
 &  & +g\psi_{j,n}\left(t\right)\omega_{q}e^{iqn}\lambda_{j,q}\\
 &  & -\frac{1}{2}\phi\psi_{j,n+1}\left(t\right)\omega_{q}e^{iqn}\left(e^{iq}-1\right)\lambda_{j,q}\left(t\right)\\
 &  & -\frac{1}{2}\phi\psi_{j,n-1}\left(t\right)\omega_{q}e^{iqn}\left(1-e^{-iq}\right)\lambda_{j,q}\left(t\right), \\ {\rm and} \\
 &   & B_{jnq} \\
 & = & ~i\psi_{j,n}\left(t\right)\dot{\lambda}_{j,q}\left(t\right)\\
 &  & -\psi_{j,n}\left(t\right)\omega_{q}\lambda_{j,q}\\
 &  & +g\psi_{j,n}\left(t\right)\omega_{q}e^{-iqn}\\
 &  & -\frac{1}{2}\phi\psi_{j,n+1}\left(t\right)\omega_{q}e^{-iqn}\left(e^{-iq}-1\right)\\
 &  & -\frac{1}{2}\phi\psi_{j,n-1}\left(t\right)\omega_{q}e^{-iqn}\left(1-e^{iq}\right).
\end{eqnarray*}

\section{Energy Conservation}

\begin{figure}[tbp]
\centering
\includegraphics[scale=0.4]{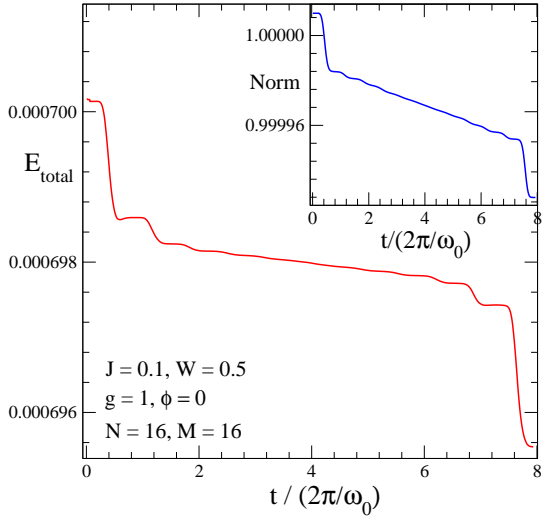}
\caption[FIG]{ In the diagonal coupling case, the total energy of the system $E_{\rm total}$ and the normalization of the wave function $ Norm$ are plotted as a function of the time $t$ for a molecular ring with $N = 16$ sites. The time unit $2\pi/\omega_0$ denotes the vibrational period of the phonon. The parameters including the transfer integral $J=0.1$, phonon energy bandwidth $W=0.5$, diagonal coupling strength $g=1$ and off-diagonal coupling strength $\phi = 0$ are set.
}
\label{long_2}
\end{figure}

In the diagonal coupling case of $J=0.1, W=0.5$ and $g=1$, the total energy $E_{\rm total}(t)$ and the normalization of the wave function $Norm(t)$ are plotted as a function of the time $t$ in Fig.~\ref{long_2} for the precision test of the multi-$\rm D_2$ {\it Ansatz} with $M=16$. One can find that the deviations of $E_{\rm total}$ and $Norm$ from the initial values are negligibly small, suggesting that the numerical results obtained by the Dirac-Frenkel variational dynamics based on the multi-$\rm D_2$ trial states are reliable.

\begin{figure}[tbp]
\centering
\includegraphics[scale=0.4]{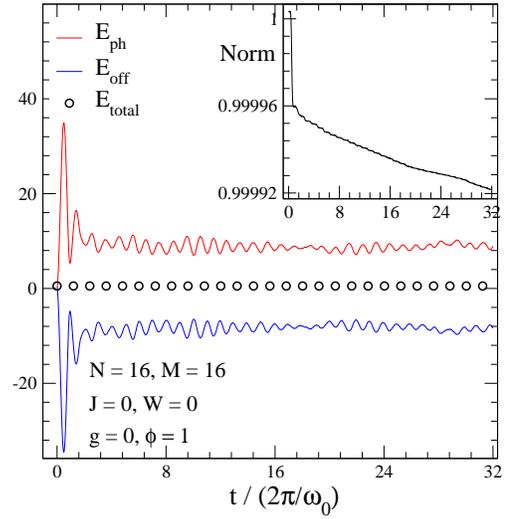}
\caption[FIG]{ In the off-diagonal coupling case with $\phi = 1$ and $J=W=g=0$, the phonon energy $E_{\rm ph}$, off-diagonal exciton-phonon interaction energy $E_{\rm off}$ and total energy $E_{\rm total}$ obtained by the multi-${\rm D}_2$ {\it Ansatz} with $M=16$, are displayed as a function of the time $t$ for a molecular ring with $N = 16$ sites. In
the inset, the normalization of the wave function $Norm$ is plotted.
}
\label{long_3}
\end{figure}

Besides, the dynamic behavior of the Holstein polaron for the off-diagional coupling case is also investigated at $J=0, W=0, g=0$ and $\phi=1$. As shown in Fig.~\ref{long_3}, aperiodic behaviors of the system energies are found, consistent with the prediction of the long period time $T\rightarrow \infty$ due to the vanishing of the band width $W=0$. $E_{\rm total}(t) = E_{\rm ph} + E_{\rm off} \approx 0$ shows the system total energy is conserved. In the inset, the normalization $Norm(t)$ is also displayed for the conservativeness test.

\section{LINEAR ABSORPTION}
Combing the Eqs.~(\ref{auto_cor}) and (\ref{polarization}), the autocorrelation function is derived
\begin{equation}
F\left(t\right)=\mu^{2}\sum_{n}\sum_{m}{}_{\rm ph}\left\langle 0\right|_{\rm ex}\left\langle 0\right|\hat{a}_{m}e^{-i\hat{H}t}\hat{a}_{n}^{\dagger}\left|0\right\rangle _{\rm ex}\left|0\right\rangle _{\rm ph}\label{eq:4}
\end{equation}
Using the periodic condition of the Hamiltonian $\hat{H}$, one has
\begin{eqnarray}
&  & \sum_{m}{}_{\rm ph}\left\langle 0\right|_{\rm ex}\left\langle 0\right|\hat{a}_{m}e^{-i\hat{H}t}\hat{a}_{n}^{\dagger}\left|0\right\rangle _{\rm ex}\left|0\right\rangle _{\rm ph}\nonumber \\
&  & =\sum_{m}{}_{\rm ph}\left\langle 0\right|_{\rm ex}\left\langle 0\right|\hat{a}_{m-n}e^{-i\hat{H}t}\hat{a}_{n-n}^{\dagger}\left|0\right\rangle _{\rm ex}\left|0\right\rangle _{\rm ph}\nonumber \\
&  & =\sum_{m}{}_{\rm ph}\left\langle 0\right|_{\rm ex}\left\langle 0\right|\hat{a}_{m}e^{-i\hat{H}t}\hat{a}_{0}^{\dagger}\left|0\right\rangle _{\rm ex}\left|0\right\rangle _{\rm ph}.
\label{eq:5}
\end{eqnarray}
Substituting Eq.~(\ref{eq:5}) into Eq.~(\ref{eq:4}), one can obtain
\begin{equation}
F\left(t\right)=\mu^{2}N\sum_{m}{}_{\rm ph}\left\langle 0\right|_{\rm ex}\left\langle 0\right|\hat{a}_{m}e^{-i\hat{H}t}\hat{a}_{0}^{\dagger}\left|0\right\rangle _{\rm ex}\left|0\right\rangle _{\rm ph},
\label{eq:6}
\end{equation}
where $e^{-i\hat{H}t}\hat{a}_{0}^{\dagger}\left|0\right\rangle _{\rm ex}\left|0\right\rangle _{\rm ph}$
is the time evolution of wave function from the initial state $\hat{a}_{0}^{\dagger}\left|0\right\rangle _{\rm ex}\left|0\right\rangle _{\rm ph}$,
which can be approximated by a Davydov trial state, for example, by the Multi-$\rm D_{2}$ trial state,
\begin{eqnarray}
 &  & e^{-i\hat{H}t}\hat{a}_{0}^{\dagger}\left|0\right\rangle _{\rm ex}\left|0\right\rangle _{\rm ph}\nonumber \\
 &  & \approx\sum_{i}^{M}\sum_{n}\psi_{i,n}\left(t\right)\hat{a}_{n}^{\dagger}\nonumber\\
 &  &  \exp\left\{ \sum_{q}\left[\lambda_{i,q}\left(t\right)\hat{b}_{q}^{\dagger}-H.c.\right]\right\} \left|0\right\rangle _{\rm ex}\left|0\right\rangle _{\rm ph}.
 \label{eq:7-1}
\end{eqnarray}
The autocorrelation $F\left(t\right)$ of the multi-$D_{2}$ {\it Ansatz} is then calculated by substituting Eq.~(\ref{eq:7-1}) into Eq.~(\ref{eq:6}),
\begin{eqnarray*}
&   &F\left(t\right) = \mu^{2}\sum_{ij}^{M}\sum_{n}\sum_{m}\psi_{i,m}\left(t\right){}_{\rm ph}\left\langle 0\right|\nonumber\\
&   &\exp\left\{ \sum_{q}\left[\lambda_{j,q}^{*}\left(t\right)\hat{b}_{q}-H.c.\right]\right\}\nonumber\\
&   &\exp\left\{ \sum_{q}\left[\lambda_{i,q}\left(t\right)\hat{b}_{q}^{\dagger}-H.c.\right]\right\} \left|0\right\rangle _{\rm ph}\\
&   &= \mu^{2}N\sum_{ij}^{M}\sum_{n}\psi_{in}\left(t\right)\nonumber\\
&   &\exp\left\{ \sum_{q}\left[-\left(\left|\lambda_{jq}\right|^{2}+\left|\lambda_{iq}\right|^{2}\right)+\lambda_{jq}^{*}\lambda_{iq}\right]\right\}
\label{M-D2}
\end{eqnarray*}

For the single $D_{2}$ trial state, the time evolution of wave function from the initial state can be approximated to
\begin{eqnarray}
 &  & e^{-i\hat{H}t}\hat{a}_{0}^{\dagger}\left|0\right\rangle _{\rm ex}\left|0\right\rangle _{\rm ph}\nonumber \\
 &  & \approx\sum_{n}\psi_{n}\left(t\right)\hat{a}_{n}^{\dagger}\nonumber \\
 &  &\exp\left\{ \sum_{q}\left[\lambda_{q}\left(t\right)\hat{b}_{q}^{\dagger}-H.c.\right]\right\} \left|0\right\rangle _{ex}\left|0\right\rangle _{\rm ph},
 \label{eq:7}
\end{eqnarray}
which leads to the autocorrelation
\begin{eqnarray*}
&  &F\left(t\right)  =  \mu^{2}\sum_{n}\sum_{m}\psi_{i,m}\left(t\right){}_{\rm ph}\left\langle 0\right| \nonumber \\
&  &\exp\left\{ \sum_{q}\left[\lambda_{q}\left(t\right)\hat{b}_{q}^{\dagger}-H.c.\right]\right\} \left|0\right\rangle _{\rm ph}\\
&  & = \mu^{2}N\sum_{m}\psi_{m}\left(t\right)\exp\left\{-\frac{1}{2}\sum_{q}\left|\lambda_{q}\left(t\right)\right|^{2}\right\}
\label{D2}
\end{eqnarray*}

Finally, the single $D_{1}$ trial state is used to calculate the time evolution of the wave function
\begin{eqnarray}
 &  & e^{-i\hat{H}t}\hat{a}_{0}^{\dagger}\left|0\right\rangle _{\rm ex}\left|0\right\rangle _{\rm ph}\nonumber \\
 &  & \approx\sum_{n}\psi_{n}\left(t\right)\hat{a}_{n}^{\dagger}\nonumber \\
 &  &\exp\left\{ \sum_{q}\left[\lambda_{nq}\left(t\right)\hat{b}_{q}^{\dagger}-H.c.\right]\right\} \left|0\right\rangle _{\rm ex}\left|0\right\rangle _{\rm ph},
\label{eq:7-2}
\end{eqnarray}
and the autocorrelation $F(t)$ is then obtained by
\begin{eqnarray*}
&  &F\left(t\right) = \mu^{2}\sum_{n}\sum_{m}\psi_{i,m}\left(t\right){}_{\rm ph}\left\langle 0\right|\nonumber \\
&  &\exp\left\{\sum_{q}\left[\lambda_{mq}\left(t\right)\hat{b}_{q}^{\dagger}-H.c.\right]\right\} \left|0\right\rangle _{\rm ph}\nonumber \\
&  & = \mu^{2}N\sum_{m}\psi_{m}\left(t\right)\exp\left(-\frac{1}{2}\sum_{q}\left|\lambda_{mq}\left(t\right)\right|^{2}\right).
\label{D1}
\end{eqnarray*}

\end{document}